\documentclass{mem}
\usepackage{natbib}\usepackage{txfonts}\usepackage{balance}
\usepackage{flushend}
\usepackage{graphicx}
\usepackage{xcolor}
\usepackage[a4paper,breaklinks,pdftex]{hyperref}
\idline{0}{0}

\begin{document}

\title{Search for ultra-light axions with the polarization of the cosmic microwave background}

\author{P. \,Diego-Palazuelos\inst{1,2}}

\institute{
Instituto de F\'isica de Cantabria (CSIC-UC), Santander, Spain
\and
Departamento de F\'isica Moderna, Universidad de Cantabria, Santander, Spain
\\
\email{diegop@ifca.unican.es}
}

\authorrunning{Diego-Palazuelos}

\titlerunning{Search for ultra-light axions with CMB polarization}

\date{Received: Day Month Year; Accepted: Day Month Year}

\abstract{
When coupled to electromagnetism via a Chern-Simons interaction, axion-like particles (ALP) produce a rotation of the plane of linear polarization of photons known as \textit{cosmic birefringence}. Recent measurements of cosmic birefringence obtained from the polarization of the cosmic microwave background (CMB) hint at the existence of an isotropic birefringence angle of $\beta\approx 0.3^\circ$, currently excluding $\beta=0$ with a statistical significance of $3.6\sigma$. Were such measurement to be confirmed as a cosmological signal, CMB information alone could constrain the ALP parameter space for masses $m_\phi\lesssim 10^{-27}$eV and axion-photon coupling constants $g_{\phi\gamma}\gtrsim 10^{-20}$GeV$^{-1}$.

\keywords{Cosmic Microwave Background -- Cosmic Birefringence -- Axion-like Particles}}

\maketitle{} 

\section{Introduction}

Axion-like particles (ALP) are one of the most popular candidates to explain dark matter and dark energy~\citep{Marsh2016}. They are described by a parity-violating pseudoscalar field, $\phi$, that can couple to the electromagnetic tensor and its dual via a Chern-Simons term in the Lagrangian density, $\mathcal{L}\subset\frac{1}{4}g_{\phi\gamma}\phi F_{\mu\nu}\tilde{F}^{\mu\nu}$. Such an interaction makes the phase velocities of right- and left-handed helicity states of photons differ, rotating the plane of linear polarization clockwise in the sky by an angle $\beta=-\frac{1}{2}g_{\phi\gamma}\int \partial\phi/\partial t dt$. This rotation is known as \textit{cosmic birefringence} because it is as if space itself behaved like a birefringent crystal~\citep{Komatsu2022}. 

Emitted at the epoch of recombination and with its polarization angular power spectra accurately predicted by the standard cosmological model, the cosmic microwave background (CMB) is the ideal tool to measure birefringence.
 
\section{A tantalising non-null $\beta$ signal}

In the past, attempts at measuring $\beta$ from CMB observations have been limited by uncertainties in the calibration of the detectors' polarization angle. Recently, a novel technique proposed by \cite{Minami2019PTEP, MinamiKomatsu2020PTEP} made possible the simultaneous determination of birefringence and polarization angles through the use of Galactic foreground emission as a calibrator. When applied to CMB data, this methodology yields a non-null isotropic birefringence angle~\citep{MinamiKomatsu2020PRL, PDP2022, Eskilt2022AA}, with the tightest constrain to date, $\beta = 0.342^{\circ+0.094^\circ}_{\phantom{\circ}-0.091^\circ}$ ($68\%$ C. L.), coming from the joint analysis of WMAP and \textit{Planck} data~\citep{EskiltKomatsu2022PRD}. 

Although robust against instrumental systematics \citep{PDP2023}, this measurement can be biased by the $EB$ correlation of Galactic dust. \cite{PDP2022} proposed two independent ways to model dust $EB$: one based on the misalignment between dust filaments and Galactic magnetic field lines~\citep{Huffenberger2020,Clark2021ApJ}, and another based on the dust templates produced in Bayesian component-separation analyses that fit parametric models to CMB data~\citep{Planck2020compsep}. 

Acknowledging the current limitations of both models, a better understanding of polarized dust emission is needed to obtain a definitive measurement of $\beta$ with this methodology. Alternatively, improving the precision of calibration techniques to accuracies of the order of $0.01^\circ$ will allow a high-significance measurement of $\beta$ without relying on Galactic foregrounds.

\section{Implications for ALP}

In a first attempt at discerning the origin of the signal, \cite{Eskilt2022AA} studied the frequency dependence of $\beta$ across the 30 through 353GHz frequency range covered by \textit{Planck}. That study concluded that the data favors a frequency-independent $\beta$ like that predicted by a Chern-Simons coupling to a pseudoscalar field while disfavoring other possible origins such as the Faraday rotation from Galactic or primordial magnetic fields. Thus, if confirmed as a cosmological signal, the $\beta\approx0.3^\circ$ found in CMB data could be attributed to an ultra-light ALP.

\begin{figure*}
\centering \resizebox{0.8\hsize}{!}{\includegraphics[clip=true]{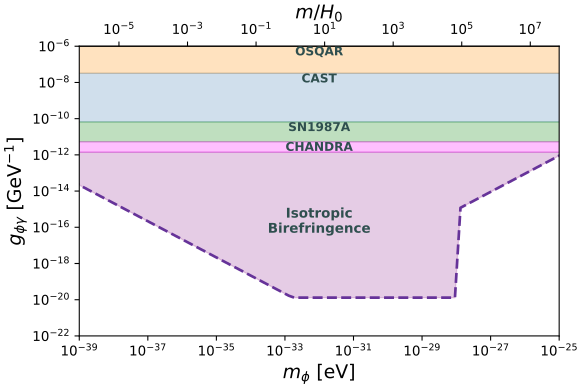}}
\caption{\footnotesize Crude constraints on the axion mass and axion-photon coupling obtained from the scaling laws derived on \cite{Fujita2021}. The shaded regions have been excluded by OSQAR~\citep{{Ballou2015}} (orange), CAST~\citep{Anastassopoulos2017} (blue), SN1987A~\citep{Payez2015} (green), and Chandra~\citep{Berg2017} (pink). Birefringence contours (purple) assume an isotropic birefringence angle of $\beta=0.30^\circ$, a tensor-to-scalar ratio of $r<0.032$, and the largest ALP abundance allowed. At the mass range considered, this last statement implies that axions would make up the entirety of dark energy.}
\label{results}
\vspace{-0.4cm}
\end{figure*}

Following the prescriptions of \cite{Fujita2021}, Figure~\ref{results} illustrates the constraining power that CMB observations alone would have on ALP parameter space if the $\beta$ measurement was confirmed. The sensitivity to the ALP-photon coupling is derived assuming a spatially flat Friedmann-Lemaître-Robertson-Walker universe and adopting a quadratic potential $V(\phi)=\frac{1}{2}m_\phi^2\phi^2$ for the ALP field. We assume the largest ALP abundance allowed~\citep{Planck2020cosmopar}, the latest constraints on the tensor-to-scalar ratio $r<0.032$~\citep{Tristram2022}, and an isotropic birefringence angle of $\beta=0.30^\circ$. 

Since the ALP abundance is only bounded from above, it is not possible to put an upper constraint on the ALP-photon coupling in Figure~\ref{results}. ALP in such a $m_\phi$-$g_{\phi\gamma}$ range could be responsible for dark energy and rule out some simple Grand Unified Theory models~\citep{Agrawal2022}.

\section{Conclusions and outlook} 

If confirmed as a cosmological signal, the measured angle of $\beta\approx 0.3^\circ$ would have profound implications for fundamental physics. It could be attributed to a Chern-Simons coupling to a light pseudoscalar field like that of ultra-light ALP~\citep{Nakatsuka2022} or early dark energy~\citep{Murai2023}. A more detailed study of the $EB$ angular power spectrum is required to distinguish between these two origins.

In addition, it would provide evidence of parity-violating physics outside the weak interaction~\citep{Lue1999}. Intriguing signs of parity violation at cosmological scales have also appeared recently in studies of galaxy spins~\citep{Motloch2022} and the four-point correlation function of galaxies~\citep{Philcox2022,Hou2022}.

To confirm this measurement, we must continue the search in independent datasets from ongoing and future CMB experiments. Improved calibration sources~\citep{Cornelison2022} and a better understanding and modeling of Galactic dust emission~\citep{Cukierman2022,Vacher2022} will be needed to obtain a definitive measurement of $\beta$.

\begin{acknowledgements}
PDP thanks the Spanish Agencia Estatal de Investigación (AEI, MICIU) for the financial support provided under project PID2019-110610RB-C21. PDP also acknowledges funding from
the \textit{Formaci\'on del Profesorado Universitario} program of the Spanish Ministerio de
Ciencia, Innovación y Universidades and the Unidad de Excelencia Mar\'ia de Maeztu (MDM-
2017-0765). 
\end{acknowledgements}


\begin{thebibliography}{}

\bibitem[Agrawal et al. (2022)]{Agrawal2022}
P. Agrawal et al., JHEP, 2022, 141 (2022)

\bibitem[Anastassopoulos et al. (2017)]{Anastassopoulos2017}
V. Anastassopoulos et al., Nature Phys., 13, 584 (2017)
 
\bibitem[Ballou et al. (2015)]{Ballou2015}
R. Ballou et al., PRD, 92, 092002 (2015)
 
\bibitem[Berg et al. (2017)]{Berg2017}
M. Berg et al., ApJ, 847, 101 (2017)
   
\bibitem[Clark et al. (2021)]{Clark2021ApJ}
S. E. Clark et al., ApJ, 919, 53 (2021)

\bibitem[Cornelison et al (2022)]{Cornelison2022}
J. Cornelison et al., Proc. SPIE 12190, 121901X (2022)

\bibitem[Cukierman et al. (2022)]{Cukierman2022}
A. J. Cukierman et al., arXiv:2208.07382 (2022)
 
\bibitem[Diego-Palazuelos et al. (2022)]{PDP2022}
P. Diego-Palazuelos et al., PRL, 128, 091302 (2022)

\bibitem[Diego-Palazuelos et al. (2023)]{PDP2023}
P. Diego-Palazuelos et al., JCAP, 01, 044 (2023)

\bibitem[Eskilt (2022)]{Eskilt2022AA}
J. R. Eskilt, A\&A, 662, A10 (2022)

\bibitem[Eskilt \& Komatsu (2022)]{EskiltKomatsu2022PRD}
J. R. Eskilt and E. Komatsu, PRD, 106, 063503 (2022)

\bibitem[Fujita et al. (2021)]{Fujita2021}
T. Fujita et al., PRD, 103, 063508 (2021)

\bibitem[Hou et al. (2022)]{Hou2022}
J. Hou et al., arXiv:2206.03625 (2022)
 
\bibitem[Huffenberger et al. (2020)]{Huffenberger2020}
K. M. Huffenberger et al., ApJ, 899, 31 (2020)
 
\bibitem[Komatsu (2022)]{Komatsu2022}
E. Komatsu, Nat. Rev. Phys., 4, 452-469 (2022)

\bibitem[Lue et al. (1999)]{Lue1999}
A. Lue et al., PRL, 83, 1506 (1999)

\bibitem[Marsh (2016)]{Marsh2016}
D. J. E. Marsh, Phys. Rep., 643, 1 (2016)

\bibitem[{Minami et al. (2019)}]{Minami2019PTEP}
Y. Minami et al., PTEP, 2019, 083E02 (2019)
 
\bibitem[Minami \& Komatsu (2020a)]{MinamiKomatsu2020PTEP}
Y. Minami and E. Komatsu, PTEP, 2020, 103E02 (2020a)

\bibitem[Minami \& Komatsu (2020b)]{MinamiKomatsu2020PRL}
Y. Minami and E. Komatsu, PRL, 125, 221301 (2020b)

\bibitem[Motloch et al. (2022)]{Motloch2022}
P. Motloch et al., PRD, 105, 083512 (2022)

\bibitem[Murai et al. (2023)]{Murai2023}
K. Murai et al., PRD, 107, L041302 (2023)

\bibitem[Nakatsuka et al. (2022)]{Nakatsuka2022}
H. Nakatsuka et al., PRD, 105, 123509 (2022)

\bibitem[Payez et al. (2015)]{Payez2015}
A. Payez et al., JCAP, 02, 006 (2015)
 
\bibitem[Philcox (2022)]{Philcox2022}
O. H. E Philcox, PRD, 106, 063501 (2022)

\bibitem[\textit{Planck} Collaboration IV (2020)]{Planck2020compsep}
\textit{Planck} Collaboration IV, A\&A, 641, A4 (2020)

\bibitem[\textit{Planck} Collaboration VI (2020)]{Planck2020cosmopar}
\textit{Planck} Collaboration VI, A\&A, 641, A6 (2020)

\bibitem[Tristram et al (2022)]{Tristram2022}
M. Tristram et al., PRD, 105, 083524 (2022)

\bibitem[Vacher et al. (2022)]{Vacher2022}
L. Vacher et al., arXiv:2210.14768 (2022)

\end{thebibliography}
\end{document}